\documentclass{emulateapj}
\usepackage{apjfonts}
\usepackage{color}

\def\eqalign#1{\null\,\vcenter{\openup\jot
        \ialign{\strut\hfil$\displaystyle{##}$&$
        \displaystyle{{}##}$\hfil \crcr#1\crcr}}\,}

\lefthead{HAN} 
\righthead{SELF-LENSING DEGENERACY}

\begin{document}
\title{Degeneracy between Lensing and Occultation in the Analysis 
of Self-lensing Phenomena}

\author{Cheongho Han}
\affil{Department of Physics, Institute for Astrophysics, Chungbuk National University, 371-763 Cheongju, 
Republic of Korea}

\begin{abstract}
More than 40 years after the first discussion, it was recently reported
the detection of a self-lensing phenomenon within a binary system where
the brightness of a background star is magnified by its foreground
companion. It is expected that the number of self-lensing binary
detections will be increased in a wealth of data from current and future
survey experiments.
In this paper, we introduce a degeneracy in the interpretation of
self-lensing light curves. The degeneracy is intrinsic to self-lensing
binaries for which both magnification by lensing and de-magnification by
occultation occur simultaneously and is caused by the difficulty in
separating the contribution of the lensing-induced magnification from the
observed light curve.
We demonstrate the severity of the degeneracy by presenting example
self-lensing light curves suffering from the degeneracy. We also present
the relation between the lensing parameters of the degenerate solutions.
The degeneracy would pose as an important obstacle in accurately determine 
self-lensing parameters and thus to characterize binaries.
\end{abstract}


\section{Introduction}

Self lensing refers to the gravitational lensing phenomenon that occurs within a binary 
system where the brightness of a background star (source) is magnified by its foreground 
companion (lens). This concept was first mentioned by \citet{Trimble1969}.
\citet{Leibovitz1971} and \citet{Maeder1973} later pointed out that binary systems in 
which one member is a degenerate, compact object -- a white dwarf (WD), neutron star (NS), 
or black hole (BH) -- could cause repeated magnification of its companion star if the 
orbit happened to be viewed edge-on. The magnification of these self-lensing binary 
systems is very tiny, typically a part in one thousand or less for a Sun-like source star, 
and thus it was considered to be very difficult to detect such 
a phenomenon.

The concept of self lensing, which had been dormant for more than two decades, resurrected 
with the advent of new types of experiments that are equipped with instruments with greatly 
enhanced photometric precision and survey capability. With the start of Galactic microlensing 
surveys, \citet{Gould1995} and later \citet{Rahvar2011} revisited self lensing and estimated 
the possibility of detecting self-lensing binaries in the data acquired from lensing surveys.  
\citet{Beskin2002} evaluated the prospect of detecting self-lensing binaries in the data 
obtained from Sloan Digital Sky Survey (SDSS).  \citet{Kasuya2011} investigated lensing effects 
of a planet transiting its host star.  \citet{Sahu2003} and \citet{Farmer2003} pointed out 
that space-based instruments such as {\it Kepler} and {\it Eddington} would be sensitive to 
compact objects in binaries through their microlensing signatures. Finally, by using {\it Kepler} 
data, \citet{Kruse2014} reported the first discovery of a self-lensing binary system (KOI-3278) 
that is composed of a WD stellar remnant and a Sun-like companion.  The number of self-lensing 
binary detections is expected to be increased in a wealth of data from current and future survey 
experiments.

Occultation and lensing are different limits of the same phenomena occurring when one 
body passes in front of another body \citep{Agol2002}.  Under some circumstances, both 
magnification of the source flux by lensing and  de-magnification by occultation can 
simultaneously occur.  This happens to binaries composed of WD-star pairs for which the 
size of the lens (WD) is equivalent to the Einstein radius \citep{Agol2002, Marsh2001}. 
For example, the radius of the WD companion of KOI-3278 is $\sim 70\%$ of the estimated 
Einstein radius.  Considering that WD-star pairs are important targets to observe 
self-lensing phenomena, there will be more self-lensing cases where both lensing and  
occultation are important.

In this paper, we introduce a degeneracy between lensing and occultation in the analysis 
of self-lensing phenomena.  We demonstrate that this degeneracy is intrinsic and thus very 
severe, making it difficult to accurately characterize self-lensing binary systems.

The paper is organized as follows.  In section 2, we describe basic physics of self lensing,
including various effects that determines self-lensing light curves.  In section 3, we 
introduce the degeneracy between lensing and occultation and demonstrate the severity of 
the degeneracy by presenting example self-lensing light curves suffering from the degeneracy.
We summarize results and conclude in section 4.

\section{Basic Physics of Self-lensing}

The basic difference between self-lensing and regular microlensing phenomena comes 
from the fact that the Einstein radius of the self-lensing phenomenon is much smaller 
than that of the regular lensing phenomenon.  The Einstein radius is related to the 
mass of the lens, $M_{\rm L}$, the distances to the lens, $D_{\rm L}$, and source, 
$D_{\rm S}$, by
\begin{equation}
r_{\rm E}=\left( {4GM_{\rm L}\over c^2}\right)^{1/2} \left[ {D_{\rm L}(D_{\rm S}-D_{\rm L})
\over D_{\rm S}}  \right]^{1/2}.
\label{eq1}
\end{equation}
For self-lensing phenomena, $D_{\rm S}\sim D_{\rm L}\gg D_{\rm S}-D_{\rm L}\sim a$ 
and thus the second term within the brackets on the right side of equation~(\ref{eq1}) 
becomes $D_{\rm L} (D_{\rm S}-D_{\rm L})/D_{\rm S}\sim a$, where $a$ is the semi-major 
axis of the binary orbit.  This term is much smaller than the factor $\sim D_{\rm L}/2$ 
of the regular lensing phenomenon that occur by a lensing object located roughly halfway 
between the source and observer and thus Einstein radii of self-lensing phenomena are 
very small.  To be also noted is that the Einstein radius of the self-lensing phenomenon 
does not depend on the distance to the binary \citep{Maeder1973}, i.e.\  
\begin{equation}
r_{\rm E}=\left({4GM_{\rm L}a\over c^2}\right)^{1/2}.
\label{eq2}
\end{equation}

Measuring the Einstein radius is important because it enables one to determine the physical 
parameters of the lens.  In addition to the relation in equation~(\ref{eq2}), the mass and 
semi-major axis are related to each other by the Kepler's law
\begin{equation}
{P^2\over a^3}={4\pi^2\over G(M_S+M_{\rm L})}.
\label{eq3}
\end{equation}
Since the orbital period $P$ and the mass of the lensed star $M_{\rm S}$ can be determined 
from follow-up photometric and spectroscopic observation, Kepler's law provides another 
relation between $M_L$ and $a$.  With the two unknowns and two relations, therefore, it is 
possible to  determine the physical parameters of self-lensing binary systems.

\begin{figure}[t]
\epsscale{1.1}
\plotone{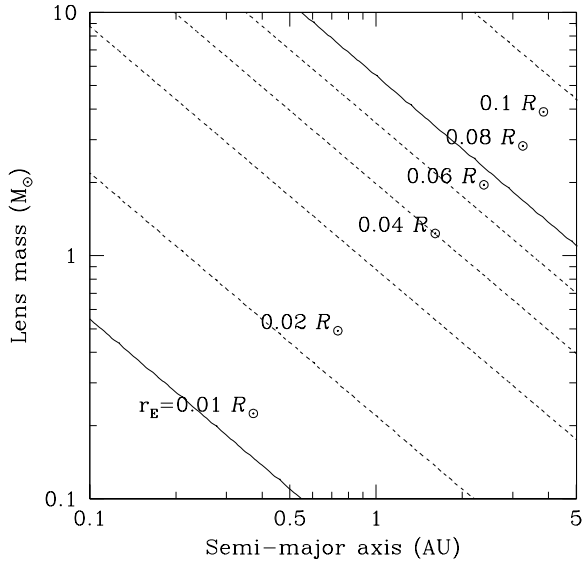}
\caption{\label{fig:one}
Einstein radii of self-lensing phenomena as a function of the semi-major axis 
of the binary and the mass of the lensing object. 
}
\end{figure}

In Figure~\ref{fig:one}, we present the Einstein radii of self-lensing phenomena as 
a function of the semi-major axis and the mass of a lensing object.  For stellar-mass 
lensing objects with masses in the range $0.1 \lesssim M/M_\odot \lesssim 10$, the expected 
Einstein radii of self-lensing effects are in the range $0.01 \lesssim r_{\rm E}/
R_\odot\lesssim 0.1$ for binaries with semi-major axes $0.1\lesssim a/{\rm AU}\lesssim 5$.  
These Einstein radii are $\sim 10^{3}$ -- $10^4$ times smaller than the Einstein radii 
of Galactic microlensing events with typical radii of $r_{\rm E}\sim (O)\ {\rm AU}$.

\subsection{Magnification by Lensing}

The most important consequence of very small Einstein radii of self-lensing phenomena
is that lensing magnifications are affected by severe finite-source effects.  Finite 
size of a source affects lensing magnifications due to a differential magnification, 
where different parts of the source surface are magnified by different amounts \citep{Witt1994}.
Finite-source effects are described by the ratio of the source radius $r_*$ to the Einstein 
radius.  For regular microlensing events, the ratio is $r_*/r_{\rm E}\sim 10^{-3}-10^{-2}$ 
and thus finite-source effects become important either for very rare cases of extremely 
high-magnification events, where the lens traverses or approaches very close to the source,
or for events associated with extremely large source stars.  On the other hand, the ratio is 
$r_*/r_{\rm E}\sim 10^{1}$ -- $10^2$ for self-lensing phenomena that occur on typical main-sequence 
stars.  As a result, self-lensing phenomena are always affected by severe finite-source effects 
and exhibit light curves that are very different from those of regular lensing events.

\begin{figure}[t]
\epsscale{1.1}
\plotone{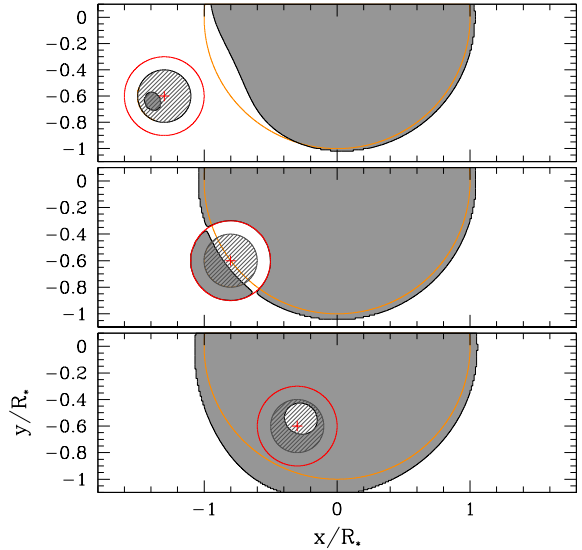}
\caption{\label{fig:two}
Variation of self-lensed images of a source with the change of the relative lens-source 
position.  The orange circle and the grey shaded region represent the source and its lensed 
image, respectively.  The lens is marked by a circle shaded by slanted lines and the ``+'' 
mark represents the center of the lens.  The red circle around the lens represents the 
Einstein ring.  The top and bottom panels show the images when the lens is out of and within 
the source surface, respectively.  The middle panel shows the image at the moment of the 
lens entrance into the source surface.  
}
\end{figure}

Under the assumption of a uniform disk, the lensing magnification of a finite source is 
computed as the area ratio of the lensed image to the source star.  The image position, 
${\bf z}$, for a given position of a source with respect to a lens, ${\bf u}$, is obtained 
by solving the lens equation
\begin{equation}
{\bf u}={\bf z}-{\bf z}^{-1},
\label{eq4}
\end{equation} 
where all lengths are normalized to the Einstein radius \citep{Paczynski1986, Witt1994}.  
Solving the equation with respect to {\bf z} yields two solutions of the image positions 
\begin{equation}
{\bf z}_{\pm}={1\over 2}\left[u \pm (u^2+4)^{1/2} \right] {{\bf u}\over u}. \\
\label{eq5}
\end{equation}

Figure~\ref{fig:two} shows the variation of self-lensing images (grey shaded region) 
with the change of the lens (circle shaded by slanted lines) location with respect 
to the source (orange circle).  We note that the positions of the image envelope 
are obtained by solving equation~(\ref{eq5}) for the positions along the envelope 
of the source.  The center of the lens is marked by the ``+'' symbol and the Einstein 
radius around the lens is marked by a red circle.  We note that a similar plot was presented 
in Figure 2 of \citet{Maeder1973}.  When the lens is out of the source star surface 
($u>r_*/r_{\rm E}$), there exist two separate images:  one big image located out of the 
Einstein ring (major image) and the other small image within the Einstein ring (minor image).  
On the other hand, when the lens is within the source star surface ($u<r_*/r_{\rm E}$), 
there exists a single image with a small hole inside.  The positions of the hole corresponds 
to those of the minor image, i.e.\  ${\bf z}_-$ in equation~(\ref{eq5}).  Then, the magnification 
of a uniform, finite source is expressed as 
\begin{equation}
A={\Sigma_{\rm I} \over \Sigma_{\rm S}}= 
\cases{
(\Sigma_{\rm I,+}+\Sigma_{\rm I,-})/\Sigma_{\rm S}  & when $u> r_*/r_{\rm E}$, \cr
(\Sigma_{\rm I,+}-\Sigma_{\rm I,-})/\Sigma_{\rm S}  & when $u< r_*/r_{\rm E}$, \cr
}
\label{eq6}
\end{equation}
where $\Sigma_{\rm S}=\pi r_*^2$ is the area of the source, and $\Sigma_{\rm I,+}$ 
and $\Sigma_{\rm I,-}$ are the areas of the major and minor images, respectively.

\begin{figure}[t]
\epsscale{1.1}
\plotone{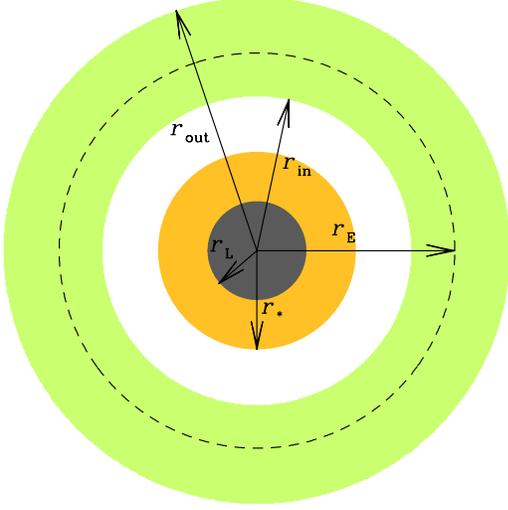}
\caption{\label{fig:three}
Geometry of a self-lensing system.  The green annulus represents the image of a source 
when the source (yellow circle) is gravitational lensed by a lens (grey circle) that 
is exactly aligned with the source.  The dashed circle is the Einstein ring.
}
\end{figure}

Under the assumption that the disk of a source is uniform, finite-source magnifications 
during very close lens-source approaches can be expressed in an analytic form.
For the derivation of the analytic expression, we present the geometry of a self-lensing 
system in Figure~\ref{fig:three}.  In the figure, the green annulus represents the image of 
a source when the source (yellow circle) is gravitational lensed by a lens (grey circle) 
that is exactly aligned with the source.  The positions of the inner and outer circles of 
the image are the major- and minor-image locations corresponding to the envelope of the 
source, and thus the radii of the inner and outer circles are
\begin{equation}
\eqalign{
r_{\rm in}  & = {1\over 2} \left[ \left( r_*^2+4r_{\rm E}^2\right)^{1/2}-r_*\right], \cr
r_{\rm out} & = {1\over 2} \left[ \left( r_*^2+4r_{\rm E}^2\right)^{1/2}+r_*\right], \cr
}
\label{eq7}
\end{equation}
respectively.  Then, the lensing magnification, which corresponds to area ratio of the 
image to the source, is expressed as \citep{Maeder1973, Riffeser2006}
\begin{equation}
A={ r_{\rm out}^2 -r_{\rm in}^2\over r_*^2}=
\left[ 1+4\left( { r_{\rm E}\over r_*}\right)^2\right]^{1/2}.
\label{eq8}
\end{equation}
In the limiting case where $r_* \gg r_{\rm E}$, equation~(\ref{eq8}) is approximated 
as\footnote{In other limiting cases, the finite magnification becomes 
$A=\sqrt{5}$ when $r_*=r_{\rm E}$ and  
$A\sim 2r_{\rm E}/r_*$   when $r_*\ll r_{\rm E}$.}
\citep{Agol2003}
\begin{equation}
A \sim 1 + 2\left( {  r_{\rm E}\over r_* }\right)^2.
\label{eq9}
\end{equation}
This approximation is useful in estimating the peak magnifications of self-lensing events.  
For example, the peak magnifications with $r_{\rm E}/r_*=0.01$, 0.05, and 0.1 are 
$A_{\rm max}=1.0002$, 1.005, and 1.02, respectively.

\begin{figure}[t]
\epsscale{1.1}
\plotone{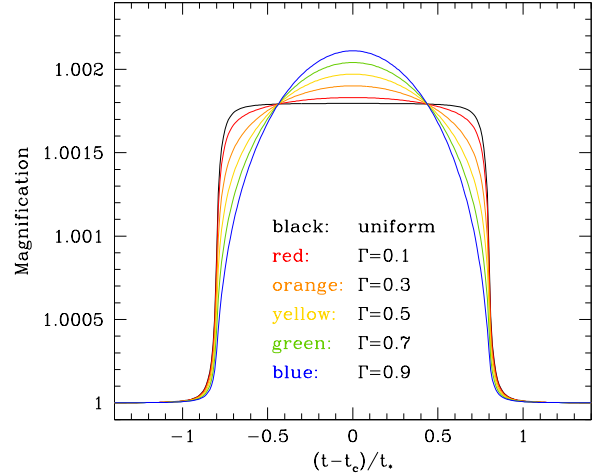}
\caption{\label{fig:four}
Variation of self-lensing light curves caused by limb-darkening effects.  The adopted radii of 
the source and Einstein ring are $r_*=1.0\ R_\odot$  and $r_{\rm E}=0.03\ R_\odot$, respectively.
The definition of the limb-darkening coefficient $\Gamma$ is provided in equation~(\ref{eq10}).
The lens-source impact parameter is $0.6\ r_*$.
The notation $t_*$ represents the time scale for the lens to cross the source radius 
and $t_{\rm c}$ represents the time of the lens approach to the center of the source.
}
\end{figure}

Stellar disks are not uniform in brightness.  Instead, central part of a stellar disk appears 
brighter than the edge due to limb-darkening.  Then, when finite-source effects are important 
in lensing magnifications, limb-darkening effects come along \citep{Witt1994}.
In Figure~\ref{fig:four}, we present the variation of lensing magnifications caused by limb-darkening 
effects.  We model the surface brightness profile as \citep{Yoo2004, Lee2009} 
\begin{equation}
S_\lambda \propto 1-\Gamma_\lambda \left(1-{3\over 2}\cos\phi \right), 
\label{eq10}
\end{equation}
where $\Gamma_\lambda$ is the linear limb-darkening coefficient and $\phi$ is the angle between 
the line of sight toward the center of the source star and the normal to the source surface.  We 
note that $\Gamma_\lambda$ depends not only on the stellar type of the source but also on the 
observed passband.  We accommodate the limb-darkening effect on lensing light curves by dividing 
the source into multiple annuli and giving a weight of the surface brightness to each annulus
in the computation of finite-source magnifications.  We note that finite-source magnifications 
including limb darkening can also be computed by using analytic expressions derived by \citet{Witt1994} 
for circular sources and \citet{Heyrovsky1997} for elliptical sources. 
From the pattern of the variation, it is found that the limb-darkened magnification is higher than 
the magnification of a uniform disk when the lens is located at the center of the source star disk, 
where the surface brightness is high.  In contrast, the magnification when the source is located 
at the edge, where the surface brightness is low, is lower than uniform-disk magnification.  It is 
also found that the limb-darkening effect causes the flat top part of the light curve to appear 
rounder.  These tendencies become more important as the limb-darkening coefficient increases.

\subsection{De-magnification by Occultation}

Very small Einstein radii of self-lensing phenomena sometimes require to consider finite sizes 
of lenses.  The effect of a finite lens on lensing magnification is caused because a lensed 
image can be partly blocked by the lens \citep{Bromley1996}.  The occultation of an image is 
illustrated in Figure~\ref{fig:two}.  Since the blocked part of the image cannot be seen, 
finite-lens effects cause the lensing magnification to appear lower than the magnification of 
a point lens.

Occultation effects occur when the lens radius is greater than the inner radius of the image, 
i.e.\ $r_{\rm L}>r_{\rm in}$.  Under this condition, lensing magnifications become \citep{Agol2003}
\begin{equation}
A={ r_{\rm out}^2 -r_{\rm L}^2\over r_*^2}
\sim 1 + 2\left( {  r_{\rm E}\over r_* }\right)^2 - \left( {r_{\rm L}\over r_*} \right)^2,
\label{eq11}
\end{equation}
where the second term on the right side is the magnification term induced by lensing while the 
last term is the de-magnification term induced by finite-lens effects.  \citet{Agol2003} showed 
that the analytic expression holds when the lens is well inside the source.  Then, the lensing 
magnification when the lens is located over the source star surface is expressed as
\begin{equation}
A \sim 
\cases{
1 + 2(r_{\rm E}/r_*)^2                       & when $r_{\rm L} < r_{\rm in}$,   \cr 
1 + 2(r_{\rm E}/r_*)^2 - (r_{\rm L}/r_*)^2   & when $r_{\rm in} < r_{\rm L} \leq r_{\rm out}$,  \cr 
0                                            & when $r_{\rm L} \geq r_{\rm out}$,   \cr 
}
\label{eq12}
\end{equation}
which is known as the ``inverse-transit approximation'' \citep{Agol2003,Marsh2001}.
Note that $A=0$ implies that the image is completely blocked out by the lens.  
We also note that the expression for exact lensing 
magnifications considering finite source and lens is provided by \citet{Lee2009}.

\begin{figure}[t]
\epsscale{1.1}
\plotone{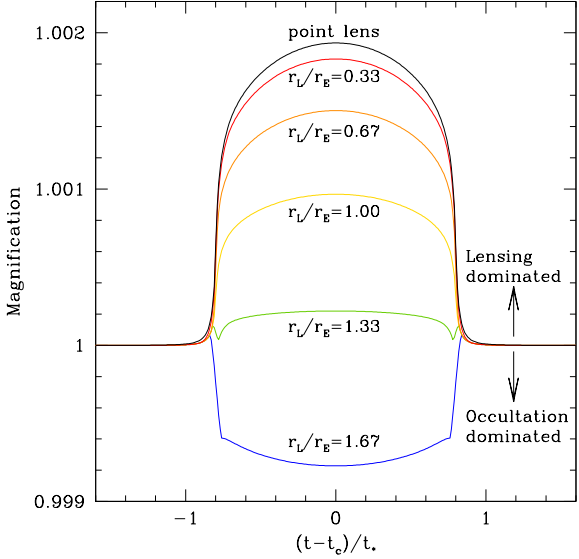}
\caption{\label{fig:five}
Variation of self-lensing light curves caused by occultation effects, which depend on ratios 
of the lens radius to the Einstein radius, $r_{\rm L}/r_{\rm E}$.  The adopted source 
radius and the limb-darkening coefficient are $r_*=1\ R_\odot$ and $\Gamma=0.4$, respectively.  
The lens-source impact parameter is $0.6r_*$.
Notations are same as those in Figure~\ref{fig:four}.
}
\end{figure}

De-magnification by occultation is important for self-lensing phenomenon where the radius of the 
lens is comparable to the Einstein radius, i.e.\ $r_{\rm E}\sim r_{\rm L}$ \citep{Lee2009}.  
For WD lenses, which 
are roughly the same size as the Earth, i.e.\ $r_{\rm L}\sim 0.01\ R_\odot$, the size of the lens 
can be a significant fraction of the Einstein radius and thus finite-lens effects can be important.
For stellar lenses, on the other hand, the lens is much bigger than the Einstein radius, 
i.e.\ $r_{\rm L} \gg r_{\rm E}$.  
In this case, the resulting light curve of the background star is simply that of an eclipsing 
binary where  lensing signatures cannot be detected.

In Figure~\ref{fig:five}, we present light curves of self-lensing events with different ratios 
of the lens radius to the Einstein radius.  From the comparison of the light curves, it is found 
that magnifying lensing effects and de-magnifying occultation effects compete each other depending 
on the ratio $r_{\rm L}/r_{\rm E}$.  For small $r_{\rm L}/r_{\rm E}$ ratios, lensing effects 
dominate and the source flux is magnified.  As the ratio increases, occultation effects become 
more important.

\begin{figure}[t]
\epsscale{1.1}
\plotone{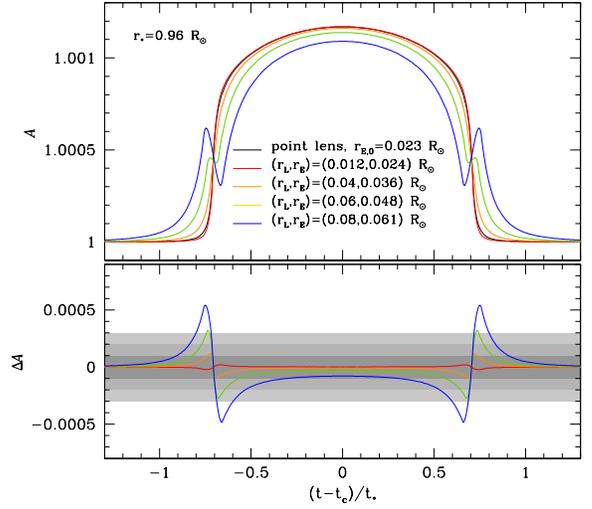}
\caption{\label{fig:six}
Self-lensing light curves with similar peak magnifications resulting from different combinations 
of the Einstein ring radius $r_{\rm E}$ and lens radius $r_{\rm L}$.  The lower panel 
shows the residual from the point-lens light curve.  The shaded regions with different grey tones 
represent the $1\sigma$, $2\sigma$, and $3\sigma$ photometric precisions of {\it Kepler} observation
achieved in 15 min sampling for a $V=12$ star.
}
\end{figure}

\section{Degeneracy}

The simultaneous occurrence of magnification by lensing and de-magnification by occultation 
provokes a question on whether  the contribution of the lensing-induced magnification can be 
separated from observed light curves in order to accurately determine the Einstein radius and 
thus to characterize binary systems.  To answer the question, we compare self-lensing light 
curves with similar peak magnifications but with different combinations of the Einstein ring 
and the lens radii.  For a given magnification and a source radius, the relation between 
$r_{\rm E}$ and $r_{\rm L}$ of degenerate solutions are obtained from equation~(\ref{eq12}), i.e.
\begin{equation}
r_{\rm E}^2={1\over 2}\left[ r_{\rm L}^2+(A-1)r_*^2 \right].
\label{eq13}
\end{equation}
We note that this expression is valid for $r_{\rm in} < r_{\rm L} \leq r_{\rm out}$.

\begin{figure}[t]
\epsscale{1.1}
\plotone{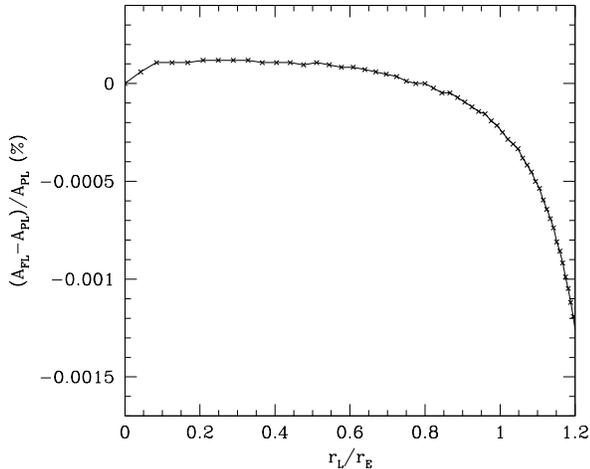}
\caption{\label{fig:seven}
Fractional deviation of the peak magnification  
affected by finite lens effects, $A_{\rm max,FL}$,
from the point-lens magnification, $A_{\rm max,PL}$, as a function of 
the lens to Einstein radius ratio, $r_{\rm L}/r_{\rm E}$.
}
\end{figure}

Figure~\ref{fig:six} shows example degenerate light curves resulting from 5 different combinations 
of $r_{\rm E}$ and $r_{\rm L}$.  The lensing parameters are chosen so that the individual light 
curves can explain the observed light curve of the self-lensing binary KOI-3278.  We note that the 
model parameters presented by \citet{Kruse2014} are  $r_*=0.96\ R_\odot$, $r_{\rm L}=0.012\ R_\odot$, 
and $r_{\rm E}=0.024\ R_\odot$ and the lens-source impact parameter is $0.706\ r_*$.  The lower panel 
shows the residuals from the point-lens case, i.e.~$r_{\rm L}=0$.  We note that for all degenerate 
cases, the lens radius is significantly larger than the inner radius of the image $r_{\rm in}$.   
In Figure~\ref{fig:seven}, we also present the distribution of the fractional deviation of the peak 
magnification $A_{\rm max,FL}$ from the point-lens magnification $A_{\rm max,PL}$ as a function of 
the lens to Einstein radius ratio, $r_{\rm L}/r_{\rm E}$.

From the comparison of the light curves, it is found that the light curves are similar each 
other despite the large differences in the values of $r_{\rm E}$ and $r_{\rm L}$.  The degeneracy 
between light curves is especially severe when the lens is smaller than the Einstein ring.
It is found that the difference in the fractional magnification is $\Delta A/A\lesssim 0.5
\times 10^{-4}$ for self-lensing events with $r_{\rm L} \lesssim r_{\rm E}$.  The photometric 
precision of {\it Kepler} is $\sim 100~\mu{\rm mag}$, 1$\sigma$ error achieved in 15 min sampling 
for a $V=12$ star \citep{Argabright2008, VanCleve2009}, which corresponds to $\Delta A/A \sim 10^{-4}$.  
Then, although self-lensing itself, with $\Delta A/A \sim 20\times 10^{-4}$, can be detected with a 
significant statistical confidence, it will be difficult to distinguish degenerate light curves 
suffering from the lensing/occultation degeneracy.

With the increase of $r_{\rm L}/r_{\rm E}$ ratio, the deviation from the point-lens light 
curve increases.  It is found that the major deviations occur during the ingress and egress of 
the lens over the source surface.  Considering the photometric precision of Kepler, however, 
it is expected that resolving the degeneracy will be possible for only events that experience 
severe occultation effects.

\section{Summary}

We introduced a degeneracy that would happen in interpreting light curves of self-lensing 
phenomena.  We found that the degeneracy was intrinsic to self-lensing binaries for which 
both magnification by lensing and de-magnification by occultation occur simultaneously.
We found that the degeneracy was severe and would be difficult to resolve by the precision 
of {\it Kepler} data.  Therefore, the degeneracy would pose as an important obstacle in 
accurately determine self-lensing parameters and thus to characterize binaries.

\acknowledgments
This Work was supported by Creative Research Initiative Program
(2009-0081561) of National Research Foundation of Korea. 
We thank A.~Gould for making useful comments.

\end{document}